\documentclass[12pt]{article}
\usepackage[dvips]{graphicx}

\def\fnote#1#2{\begingroup\def\thefootnote{#1}\footnote{#2}
    \addtocounter{footnote}{-1}\endgroup}

\newcommand{\be}{\begin{equation}}
\newcommand{\ee}{\end{equation}}
\newcommand{\bea}{\begin{eqnarray}}
\newcommand{\eea}{\end{eqnarray}}

\def\abstract#1{\begin{center}{\large ABSTRACT}\end{center} \par #1}
\def\title#1{\begin{center}{\Large\bf {#1}}\end{center}}
\def\author#1{\begin{center}{\large #1}\end{center}}
\def\address#1{\begin{center}{\it #1}\end{center}}

\begin{document}
\begin{titlepage}
\hspace*{\fill}
\vbox{
\hbox{UT-Komaba/00-06}
\hbox{hep-th/0003241}
  \hbox{March, 2000}}
\vspace*{\fill}
\begin{center}
  \Large\bf
Compactification and Identification of \\
Branes in the Kaluza-Klein monopole backgrounds
\end{center}
\vskip 1cm
\author{
Masako Asano \fnote{\dag}{ 
E-mail address: {\tt asano@hep1.c.u-tokyo.ac.jp};
Address after April 1, 2000: Department of Physics, Faculty of Science,
University of Tokyo, Tokyo 113-0033, Japan.}}
\address{
Institute of Physics, \\
  University of Tokyo, \\
  Komaba, Meguro, Tokyo 153-8902, Japan}
\vspace{\fill}
\abstract{
We study the properties of branes in supergravity theory. 
We investigate a class of systems consisting of an M5-brane in the Kaluza-Klein
monopole background with 1/4 supersymmetry in 11-dimensions.
In the near core region of the KK-monopoles, 
the exact supergravity solution corresponding to 
each of these configurations is obtained.
Then we argue the compactified 10-dimensional systems and 
suggest a way of unambiguous identification of 
branes in this background.
Here the location of Dirac string type singularity accompanied by the D6-branes
plays an important role.
The method is essentially the same as that of 
$(p,q)$5-branes or $(p,q)$-strings within the 7-brane background
in the IIB theory.
We also argue the phenomena of D4-brane creation from D6-branes.
} \vspace*{\fill}


\end{titlepage}

\section{Introduction} 
It has been well-known that there could be various types of extended objects 
in string theory and 11-dimensional M-theory.
D-branes~\cite{P} in superstring theories made us possible to study the
duality relations among different theories.
In particular, duality relation between gauge theory and supergravity or 
superstring theory, particularly AdS/CFT correspondence \cite{Mal},
has been investigated by using the D-branes as intermediates.
Moreover, it has been discussed so much that moduli space structure of
gauge theory can be read from the various brane configurations in background 
of superstring theory or M-theory \cite{HW,W}.
Thus, it is important to study the properties of branes in various backgrounds  
in order to deepen the understanding of gauge theory and string theory
in the above viewpoints.

Branes in 10-dimensional type IIA string theory originate from 11-dimensional M-theory
at least in the sense of low energy effective supergravity theory \cite{To}.
For example, D4-branes and solitonic 5-branes (NS5-branes) in IIA supergravity theory 
are obtained by dimensional reduction of solitonic M5-branes, and fundamental strings
and D2-branes are from membranes.
As for D6-branes which carry Kaluza-Klein magnetic charges, original 11-dimensional
counterpart is Kaluza-Klein monopole solution \cite{SO,GP,GH}.
It is described as the 11-dimensional background which is direct product of 7-dimensional
Minkowski space ${\cal M}^{1,6}$ and the Euclidean Taub-NUT space $M_{TN}$.

We can further consider stable M-theory backgrounds where several different 
types of branes live together.
A family of configurations representing process of brane creation from another brane 
belongs to this class \cite{HW,W,GZ,NOYY}.
However, since little is known about the explicit supergravity solutions corresponding to
such complicated configurations, the process of brane creation in terms of supergravity 
is still not understood well enough.

Now, consider a supersymmetric configuration of M-theory which has $N$ coincident
KK-monopoles and an M5-brane of world-volume $ R^{1,3}\times \Sigma $
with $\Sigma \subset M_{TN}$.
Upon compactification to 10-dimensional IIA theory, this becomes a configuration of an
NS5-brane, a D4-brane or some bound state of these two 
in the background of $N$ D6-branes.
Of such a class of systems, there is a family of configurations representing
brane creation phenomenon.
In a flat spacetime background, it is represented as the following process:
If an NS5-brane of world-volume  
$\{x_0, \cdots , x_3, x_7,x_8 \}$ crosses a D6-brane of world-volume 
$\{x_0, \cdots , x_3, x_4, x_5, x_6 \}$,
a new finite D4-brane between these two branes is created.
Here $x_i$ $(i=0,\cdots ,9)$ denote the spacetime coordinates.
This process is U-dual to the original Hanany-Witten configuration \cite{HW}.
An attempt of investigating such a phenomenon in 11-dimensional 
viewpoint was done in e.g., refs.\cite{W,NOYY}.
In particular, in ref.\cite{NOYY},
one parameter family of M5-branes in the supergravity
background of a KK-monopole and its compactified counterpart is investigated.

On the other hand, in another context, explicit supergravity solutions describing 
M2- or M5-branes localized near core of KK-monopoles
were constructed by using the fact that the metric becomes flat 
in the vicinity of KK-monopoles \cite{ITY,AH}.
The solution of an M5-brane with world-volume 
$R^{1,3} \times \Sigma_0$ ($\Sigma_0 \subset M_{TN}$)
near core of KK-monopoles was found in ref.\cite{AH}.
It is naturally expected that the near core version of brane creation phenomenon explained 
in the last paragraph can be described by dimensional reduction of this solution.
In practice, in refs.\cite{AH,GM}, some discussion about brane creation 
in 10-dimensional viewpoint was done.

It was pointed out in ref.\cite{AH} that in the near core region of KK-monopoles, 
there is some ambiguity in the definition of Ramond-Ramond four-form field
strength and NS-NS three-form field strength.
This means that the identification of D4-branes and NS5-branes cannot be done exactly.
In ref.\cite{GM}, this problem was further studied and 
a resolution by introducing a certain {\it non-conserved} charge was suggested.
Using this procedure, it was argued there that an NS5-brane 
is transmuted into a D4-brane in a certain limit and that this phenomenon is a near
core version of the brane creation.
There still remains a problem in the sense that the argument is restricted in the
near core region of KK-monopoles and that the role of non-conserved charge is unclear.

Our aim of the present paper is to confirm the definition of brane current ({\it i.e.},
the identification of branes) and to investigate 
the brane creation phenomena in 10-dimensions for systems
of an M5-brane in the background of $N$ KK-monopoles.
As for the definition of current associated with branes, 
we adopt the natural definition of current which assigns {\it conserved}  
charge to branes in the background of D6-branes.
The ambiguity presented above can be resolved by 
noticing that the location of Dirac string type singularity coming up from the
D6-branes is relevant to the identification of D4-branes.
This interpretation is essentially the same as the charge assignment of 
the systems defined by T-dual
of $(p,q)$5-branes or $(p,q)$-strings \cite{Sc}
in the 7-brane background \cite{GZ,GSVY,Va,MA}.

We also give an explicit relation between the complex structures
of $M_{TN}$ and some specific complex coordinates of the near core flat space.
As a result, it becomes possible to discuss the brane creation process 
of ref.\cite{NOYY} in terms of exact
supergravity solution in the near core region of KK-monopoles.
By applying our definition of brane charge and the way of identifying branes, we will
argue that D4-branes seem to come up from the Dirac string type
singularity and go along the NS5-brane.
If we take a limit such that net NS5-brane charge disappears,
it can be seen that only D4-branes come up from D6-branes.

\smallskip

The organization of this paper is as follows.
We begin in section~2 to review the Kaluza-Klein monopole solutions in 11-dimensional
background. 
We explain the properties of Euclidean Taub-NUT space $M_{TN}$ and 
explicitly give three independent complex structures 
by defining holomorphic coordinates corresponding to them.
In section~3 we see that in the near core region of KK-monopoles, 
the background $M_{TN}$ reduces to the flat space.
Moreover, we specify the complex coordinates of this flat space 
so that they are connected to the complex structures of $M_{TN}$.
In section~4 we introduce a family of complex one-dimensional curves 
$\{\Sigma \}$ in $M_{TN}$ which respectively correspond to configurations of an M5-brane of world-volume $R^{1,3}\times \Sigma $ in the KK-monopole backgrounds.
Each of the curves are taken to be holomorphic with respect to one of the 
complex structures.
In section~5, we further study such systems by restricting 
in the near core region of KK-monopoles and give their supergravity solutions. 
In section~6, we perform compactification of the systems 
and explain the ambiguity concerning the identification of D4-branes.
In section~7 we digress from our systems for a moment and 
argue the configuration of M-branes in the stringy cosmic string 
background and its compactification.
In particular, we see that the Dirac string type singularity, which appears as 
IIA counterpart of the branch cut related to SL(2,{\bf Z}) invariance 
of type IIB theory,
has an important role in identifying the type of branes.
Then we return to the KK-monopole backgrounds in senction~8 and by using 
the analogy with the way of identifying branes in the stringy cosmic string 
background, we give an unambiguous identification of D4-branes.
In section~9, using our definition of brane currents,
we investigate the phenomena of brane creation.
In the final section~10, we conclude with some discussion.
\section{Kaluza-Klein monopoles and the Taub-NUT space}
We now review the 11-dimensional supergravity solution of
$N$ coincident Kaluza-Klein monopoles which represents $N$ D6-branes 
after compactifying to 10-dimensions.
The solution is given by taking the 11-dimensional spacetime as a product of 
7-dimensional Minkowski space ${\cal M}^{1,6}$ and the four dimensional
 Euclidean multi-centered Taub-NUT space $M_{TN}$ \cite{SO,GP,GH}. 
The Taub-Nut space is a Hyper-K\"ahler manifold which admits three 
independent complex structures.

For the Taub-NUT space with $A_{N-1}$ singularity at $r=0$,
the metric is given by
\be
ds_{TN}^2 = V\, d\vec{r} \cdot d\vec{r} + V^{-1} (dx_{11} + \vec{A}\cdot 
d\vec{r}\,)^2
\label{eq:TN}
\ee
where $x_{11}$ is a compact direction whose radius is $R$ :
$x_{11}\sim x_{11} + 2\pi R$, and
\be
\vec{r}=(r_1,r_2,r_3),\quad
V=1+\frac{N\! R }{2r}, \quad  \vec{\nabla}\times 
\vec{A} = \vec{\nabla}\, V .
\ee
We explicitly choose $\vec{A}$ to be
\be
\vec{A}\cdot d\vec{r}=\frac{N\! R}{2}(\cos\theta - 1) d\psi
\label{eq:Ax}
\ee
where 
\be
r_1 = r \cos\theta, \; r_2= r \sin\theta \cos\psi, \; r_3= r \sin\theta \sin\psi  
\ee
and $0 \leq \psi \leq 2\pi $. 
Since the space is hyper-K\"ahler and has Ricci-flat metric, 
it solves Einstein equations and admits half of supersymmetry.
We will only deal with the bosonic part of the theory.

The metric eq.(\ref{eq:TN}) with eq.(\ref{eq:Ax})
has a coordinate singularity at $\theta = \pi$, {\it i.e.},
the negative $r_1$-axis.  
After dimensional reduction, seen in terms of ten dimensions,
this singularity is identified as the Dirac string singularity
coming up from the D6-branes.
This singularity can be moved to another place by carrying out a
coordinate transformation.
The simplest example is the transformation 
$(\vec{r},x_{11})\rightarrow (\vec{r},y_{11})$ where $y_{11}\equiv x_{11}- N\! R \psi$.
Then, by the relation 
$d x_{11}+\vec{A}\cdot d\vec{r}=d y_{11}+\vec{A'}\cdot d\vec{r}$ with
\be
\vec{A'}\cdot d\vec{r} = \frac{N\! R}{2}(\cos\theta +1) d\psi,
\ee
the coordinate singularity is moved to the positive $r_1$-axis.

Now we give explicitly three independent complex structures of the space $M_{TN}$.
First we present a natural choice of complex structure for the coordinate system
$(\vec{r},x_{11})$.
It is specified by the holomorphic complex variables $(v_0,w_0)$ as \cite{NOYY}
\bea
v_0 &=& \frac 1{N\! R}(r_2+ i r_3),
\label{eq:v0}
\\
w_0 &=& \sqrt{\frac{r+r_1}{N\! R}} \; 
\exp\left(\frac 1 {N\! R}(r_1+ i x_{11})\right).
\label{eq:w0}
\eea
Here we take $v_0$ and $w_0$ to be dimensionless.
Remembering $x_{11} \sim x_{11} +2\pi R $, $w_0^N$ may be more useful than $w_0$ as a coordinate variable.
Other two complex structures that are orthogonal to the above one
are represented by the complex variables
$(v_1,w_1)$ and $(v_2,w_2)$ 
which are holomorphic with respect to the 
complex structures respectively:
\be
\left\{
\begin{array}{rl}
v_1 &= \frac 1{N\! R}(r_1+ i r_2),
\\
w_1 &=  \sqrt{\frac{r+r_3}{N\! R}} \; 
\exp\left(\frac 1 {N\! R}(r_3+ i x_{11}^{(1)})\right)
\end{array}
\right.
\ee
and
\be
\left\{
\begin{array}{rl}
v_2 &= \frac 1{N\! R}(r_3+ i r_1),
\\
w_2 &= \sqrt{\frac{r+r_2}{N\! R}} \; 
\exp\left(\frac 1 {N\! R}(r_2+ i x_{11}^{(2)})\right) .
\end{array}
\right.
\ee
Here $x_{11}^{(1)}$ and $x_{11}^{(2)}$ are compact coordinates 
with period $2\pi R$ determined by the differential equations
\be
dx_{11}^{(1)}= dx_{11}-\frac{N\! R}{2r}
\left(
\frac{r_2dr_3-r_3dr_2}{r+r_1}-\frac{r_1dr_2-r_2dr_1}{r+r_3} 
\right)(=dx_{11}- N\! R\, d\chi)
\label{eq:dx111}
\ee
and 
\be
dx_{11}^{(2)}= dx_{11}-\frac{N\! R}{2r}
\left(
\frac{r_2dr_3-r_3dr_2}{r+r_1}-\frac{r_3dr_1-r_1dr_3}{r+r_2}
\right).
\ee
These equations can be integrated explicitly except on
singularity: negative $r_1$ and $r_3$ axes for $x_{11}^{(1)}$, and 
negative $r_1$ and $r_2$ axes for $x_{11}^{(2)}$.
These singularities are due to the coordinate transformation from $(\vec{r},x_{11})$
(which is singular on negative $r_1$ axis)
to $(\vec{r},x_{11}^{(1)})$ 
(which is singular on negative $r_3$ axis) or to $(\vec{r},x_{11}^{(2)})$.

Note that the metric in the coordinate system $(\vec{r},x_{11}^{(1)})$ is 
\be
ds_{TN}^2 = V d\vec{r} \cdot d\vec{r} + V^{-1} 
(dx_{11}^{(1)} +  \vec{A''}\cdot d\vec{r})^2
\ee
where 
\be
\vec{A''}\cdot d\vec{r}= -\frac{N\! R}{2r} \;
\frac{r_1dr_2-r_2dr_1}{r+r_3}.
\ee
If we compactify along the $x_{11}^{(1)}$ direction 
instead of $x_{11}$ direction,
then the resulting 10-dimensional D6-brane solution has 
Dirac string singularity on negative $r_3$ axis.
The movement of the singularity corresponds to  
gauge transformation: $\vec{A}\rightarrow \vec{A''}=\vec{A}+ d\chi$.
In the following discussion, we fix the compactification direction
as along $x_{11}$.

\section{Near core region of KK-monopoles}
In this section, we see that the metric $ds _{TN}^2$ reduces to the flat metric in the
near core region of KK-monopoles \cite{IMSY}. 
We also show that a flat complex coordinate system in this region is exactly related to 
one of the complex structures described in section~2.
Note that by the near core region, we mean
the region $r \ll N\! R$ in terms of the coordinate system $(\vec{r},x_{11})$.

In the limit $r/N\! R  \rightarrow 0$, 
the metric (\ref{eq:TN}) becomes 
\be
ds_{r\rightarrow 0}^2 =  \frac{N\! R}{2r} d\vec{r} \cdot d\vec{r} 
   + \frac{2r}{N\! R } \left(dx_{11} + \frac{N\! R}{2}(\cos\theta - 1) d\psi\right)^2 
\label{eq:TNr03}
\ee
since $V(=1+\frac{N\! R}{2r})\rightarrow  \frac{N\! R}{2r}$.
By changing variables from $(r, \theta, \psi, x_{11})$ to
$(\rho, \tilde{\theta},\tilde{\psi},\tilde{\phi})$ such as
\be
\rho=\sqrt{2 N\! R r},\quad \tilde{\theta}=\frac{\theta} 2 ,\quad
\tilde{\psi}=-\psi + \frac{x_{11}}{N\! R} ,\quad \tilde{\phi}=\frac{x_{11}}{N\! R},
\ee
the metric (\ref{eq:TNr03}) can be rewritten:
\be
ds_{r\rightarrow 0}^2 = d\rho^2+\rho^2d\tilde{\theta}^2
              + \rho^2(\sin^2\tilde{\theta}\, d\tilde{\psi}^2 
                    + \cos^2\tilde{\theta}\, d\tilde{\phi}^2) .
\label{eq:TNr02}
\ee
Here the range of the variables is
\be
\rho \geq 0,\quad 0\leq \tilde{\theta}\leq \frac {\pi}{2},
\quad 0 \leq \tilde{\phi},\,\tilde{\psi} \leq 2\pi
\ee
with the ${\bf {Z}}_{N}$ identification 
$(\tilde{\phi},\tilde{\psi})\sim (\tilde{\phi},\,\tilde{\psi})+(2\pi/N,2\pi/N)$.

Furthermore, defining the complex variables as%
\footnote{Here we choose the complex variables $V$ and $W$
such that they are connected to holomorphic variables
of the complex structure $(v_0,w_0)$. 
}
\be
W= \rho e^{i \tilde{\phi}} \cos\tilde{\theta},\quad
V= \rho e^{-i \tilde{\psi}} \sin\tilde{\theta},
\ee 
eq.(\ref{eq:TNr02}) becomes
\be
ds_{r\rightarrow 0}^2 = dW d\overline{W} +dV d\overline{V} 
\ee
where $(W,V)\sim (W\,e^{2\pi i/N}, V\,e^{-2\pi i/N})$.
Thus we see that the space is a flat orbifold ${\bf C}^2/{\bf Z}_N$
which is considered as
the ALE space with $A_{N-1}$ singularity at the origin.
For later convenience we represent $(W,V)$ by the original variables:
\bea
W&=& \sqrt{N\! R}\,e^{i\frac{x_{1\! 1}}{N\! R}}\sqrt{r+r_1},\\
V&=& \sqrt{N\! R}\,e^{-i\frac{x_{1\! 1}}{N\! R}}\frac{r_2+i\,r_3}{\sqrt{r+r_1}}
\nonumber\\
&=& N\! R\;W^{-1} \,(r_2+ i r_3).
\eea
Note that there is no coordinate singularity in $(W,V)$ space
except at the orbifold point.
The coordinate singularity $\theta =\pi$ in the $(\vec{r},x_{11})$
coordinate system corresponds to $W=0$, which is completely smooth in the 
new coordinate system.
Similarly, $\theta=0$, which corresponds to coordinate singularity in the 
$(\vec{r},y_{11})$ system, appears as $V=0$ which is regular in 
$(W,V)$ space.

Next, we study the relation between $(W,V)$ and the complex structure of $M_{TN}$
represented by $(w_0,v_0)$.
The behavior of $w_0$ and $v_0$ in the near core limit is obtained 
by extracting the lowest order terms of $r_i/N\! R$ in the 
eqs.(\ref{eq:v0}) and (\ref{eq:w0}):
\bea
w_0 &\rightarrow& \sqrt{\frac{r+r_1}{N\! R}} e^{i\frac{x_{1\! 1}}{N\! R}}
 \quad(\equiv w_{0(r\rightarrow 0)}),
\label{eq:w0w}
\\
v_0 &=& \frac 1{N\! R}(r_2+ i r_3) \quad (\equiv v_{0(r\rightarrow 0)}).
\label{eq:v0v}
\eea
Thus, we see that the complex variables $(W,V)$
are related to the $(w_0,v_0)$ of $M_{TN}$ as
\bea
w_{0(r\rightarrow 0)} &=& \frac{1}{N\! R} W ,
\label{eq:w0w2}\\
v_{0(r\rightarrow 0)} &=& \frac{1}{(N\! R)^2}\, V\,W  .
\label{eq:v0v2}
\eea
This means that 
in the near core limit, we can take $(W,V)$ as variables representing 
the complex structure.

We can also relate the variables $(W,V)$ with 
other complex structures of $M_{TN}$.
In the near core flat space, 
a complex structure orthogonal to $(W,V)$ is given by
a linear combination of two complex structures :
\be
(z_7+ iz_8, z_9 + i z_{10})\quad {\rm and}\quad
 (z_9+ iz_7, z_8 + i z_{10})
\label{eq:nc12}
\ee
where we introduced real variables $z_i$ as
\be
W= z_7+i z_{10},\quad V=z_8+ iz_9.
\ee
In practice, 
we are able to show as the same way as the above discussion
that two complex structures presented in eq.(\ref{eq:nc12})
are considered as the limiting representations of
the complex structures $(w_1,v_1)$ and $(w_2,v_2)$ 
of $M_{TN}$ respectively.
For example, in the $r/N\! R \rightarrow 0$ limit,
the variables $(w_1,v_1)$ become
\bea
w_{1(r \rightarrow 0)} &=& \frac{1}{N\! R} W^{(1)} ,
\\
v_{1} &=& \frac{1}{N\! R} W^{(1)}V^{(1)}
\eea
where
\bea
W^{(1)}& =&  \frac{1}{\sqrt{2}} ((z_7+ iz_8)+(z_9 + i z_{10})),
\\
V^{(1)}&=& \frac{1}{\sqrt{2}} ((z_7+ iz_8)-(z_9 + i z_{10})).
\eea
We again have a relation $(W^{(1)},V^{(1)})\sim 
(W^{(1)}\,e^{2\pi i/N}, V^{(1)}\,e^{-2\pi i/N})$.
Note that as for a choice of origin of the two periodic coordinates 
$x_{11}$ and $x_{11}^{(1)}$ in the flat $(W,V)$ space, we have taken 
$x_{11}\equiv N\! R \arg(z_7+ iz_{10})$
and $x_{11}^{(1)}\equiv N\! R \arg(W^{(1)}+V^{(1)})$.

\section{M-branes in the KK-monopole background}
We consider to put an M5-brane or an M2-brane in the 
KK-monopole background ${\cal M}^{1,6} \times M_{TN}$ such that 
two spatial dimensions of the M-brane are included in $M_{TN}$,
{\it i.e.}, the world volume of the M5-brane (or M2-brane) is
$R^{1,3}\times \Sigma$ (or $R^{1,0}\times \Sigma$)
where $\Sigma$ is a two-dimensional surface contained in $M_{TN}$.  
By the discussion in refs.~\cite{W,BBS}, 
the curve $\Sigma\subset M_{TN}$ must be holomorphic 
with respect to a complex structure 
in order to preserve a certain number, in this case $1/4$ of 32,
 of supersymmetry.
{}From now on, we concentrate on the case of M5-branes.

Now we consider two particular curves in $M_{TN}$
discussed in refs.~\cite{NOYY,BG} 
which are holomorphic with respect to $w_0$ and $v_0$:
 \bea
(A) && (w_0)^N  = e^{-\frac{ b +i\alpha}{R}} ,
\label{eq:wv0a}
\\
(B)&& (w_0)^N  = e^{\frac {b + i\alpha} {R}} \,(v_0)^N
\label{eq:wv0b}
\eea
where $b$ and $\alpha $ are some real parameters.

\begin{figure}[hbt]
    \begin{center}
\includegraphics[width=16.5cm,clip]{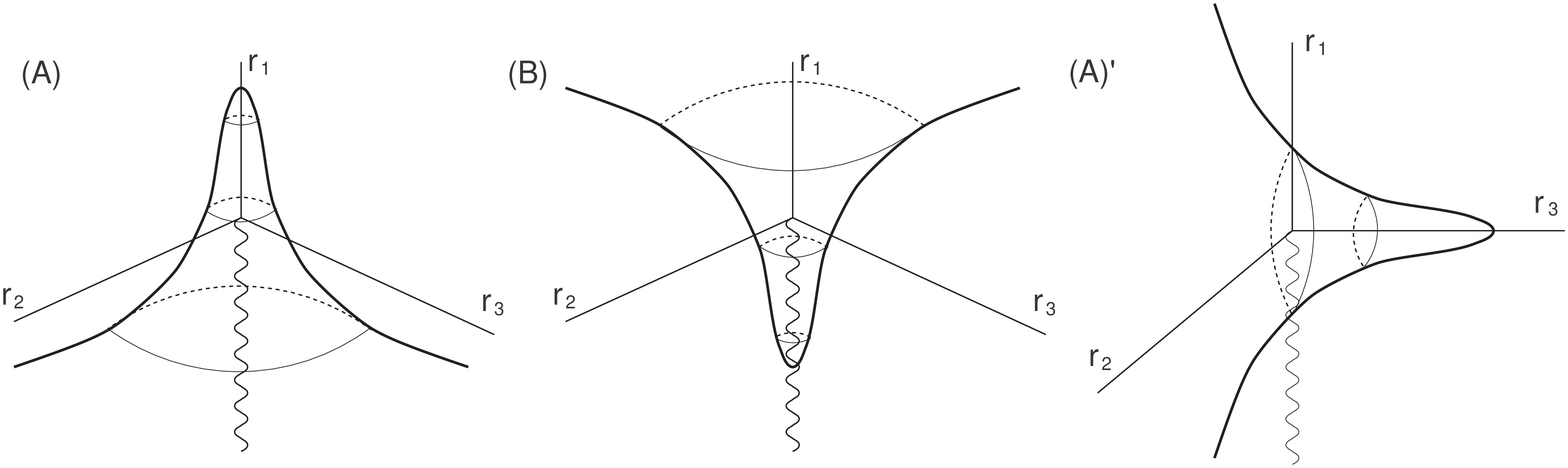}
    \end{center}
    \caption{\small Three representative curves $(A)$, $(B)$ and $(A)'$
are depicted in $(r_1,r_2,r_3)$ space.
Wavy lines represent the string like coordinate singularity.
The values of $x_{11}$ are given respectively as $(A)$ $x_{11}= - \alpha$,
$(B)$ $x_{11}=N\! R\psi +\alpha$ and $(A)'$ $x_{11}= N\! R \chi - \alpha $.
}
    \label{fig:curveform}
  \end{figure}

Remember that $M_{TN}$ is spanned by the coordinate system
$(\vec{r},x_{11})$ except on the negative $r_1$ axis.
By representing above two curves in this coordinate system,
we can investigate the shape of curves as embedding 
objects in $(r_1,r_2,r_3)$ and the behavior of the parameter 
$x_{11}$ on these curves.
It can easily be seen that the two curves with the same $b$
have exactly the same shape in $(r_1,r_2,r_3)$.
The curve $(A)$ is 
\be
\sqrt{\frac{r+r_1}{N\! R}}=e^{-\frac{r_1+b}{N\! R}}
\label{eq:formr}
\ee
and the curve $(B)$ is represented as the same equation but 
with $r_1 \rightarrow -r_1$.
The behavior of the parameter $x_{11}\! =\! x_{11}(\vec{r})$ is as follows:
The curve $(A)$ is at $x_{11}\! =\! - \alpha $, while $(B)$ 
is winding around $x_{11}$ along the $\psi$ [$=\arg (r_2+i r_3)$] direction as
$ x_{11}\! =\! N\! R \psi+ \alpha $.
The important point is that the locations of coordinate singularity, 
which will be interpreted as the Dirac string singularity in 10-dimendions,
relative to the curves $(A)$ and $(B)$ are different from each other.
(See Fig.\ref{fig:curveform}.)

In the case of $N\! =\! 1$,
it was argued in refs.\cite{NOYY,BG} that
each of these two curves corresponds to 
a single NS5-brane or a configuration 
consisting of a finite D4-brane between an NS5-brane and a D6-brane
in the compactified 10-dimensions.
We postpone the analysis of compactification until section 6.
We notice that there are many other possible curves which leave the same 
number of supersymmetry and have the same shape as the above two curves:
We can take the curves holomorphic to any linear combination of
all three complex structures of $M_{TN}$.
They all have the same shape in $(r_1,r_2,r_3)$ space 
and are represented by the equations rotating eq.(\ref{eq:formr}) 
around the $r=0$ point suitably.
The location of the singularity is still on the negative $r_1$ axis.
The value of $x_{11}$ on each of the curves 
is represented as $x_{11}=f(\vec{r})-\alpha$.
The form of the function $f=f(\vec{r})$ is generically not a simple form 
except for the curve based on the complex structure $(w_0,v_0)$.
For example, the curves holomorphic with respect to $(v_1,w_1)$ are given as
\bea
(A) '&& (w_1)^N  = e^{-\frac {b + i\alpha} {R}} ,
\label{eq:a1nc}
\\
(B) '&& (w_1)^N  = e^{\frac {b + i\alpha} {R}} \,(v_1)^N .
\label{eq:b1nc}
\eea
For the curve $(A)'$, $f(\vec{r})=N\! R \chi $
 where $\chi $ is given by eq.(\ref{eq:dx111}).

In the following discussion, we only deal with a class of M5-branes specified by 
the curves we have given above.
In particular, we often proceed with discussion by choosing the curves
$(A)$, $(B)$ and $(A)'$ as examples.

\section{M5-branes in the near core region of KK-monopoles and the 
supergravity solutions}
In the last section, we considered a family of curves which correspond to configurations of
an M5-brane in the background of KK-monopoles.
Supergravity solutions representing such systems have not been constructed. 
However, if we restrict ourselves to the near core region of KK-monopoles, 
solutions can be obtained by using the method of \cite{ITY,AH}.
We know that the near core region of KK-monopole solution is represented
by the flat complex coordinates $(W,V)$.
Thus, by considering some two-dimensional flat plane $\Sigma_0$ 
in this region, supergravity solution of an M5-brane with world-volume 
$R^{1,3}\times \Sigma_0$ is obtained.
By transforming back the coordinates to $(\vec{r},x_{11})$,
we know the behavior of the M5-brane 
in the near core region of KK-monopoles~\cite{AH}.

Now, from the analysis in section~3, we have the relation between 
the near core coordinates $(W,V)$ and the complex structures of the whole Taub-NUT
space. 
Using this knowledge, we study the behavior of the class of holomorphic curves
considered in the last section in the near core region.
In order to do this, parts of the curves must be in the near core region,
{\it i.e.}, the distance between the curve and the core of 
the monopoles $r=0$ has to be much smaller than $N\! R$.
This is equivalent to the condition $e^{-b/N\! R} \ll 1$.
Assuming this,
by the relations (\ref{eq:w0w2}) and (\ref{eq:v0v2}),
the two curves $(A)$ and $(B)$ given in eqs. (\ref{eq:wv0a}) and (\ref{eq:wv0b})
can be approximated in the $r/N\! R \rightarrow 0$ limit as
\bea
\tilde{(A)}  &&  W^N  =  c^N,
\\
\tilde{(B)} &&  V^N  =  c^N
\eea
where 
\be
c= N\! R \, e^{-\frac{b + i \alpha}{N\! R}} .
\ee
Also, the curves $(A)'$ and $(B)'$ in (\ref{eq:a1nc}) and (\ref{eq:b1nc})
can be rewritten in this region as
\bea
\tilde{(A)'} && ( W^{(1)}) ^N  =c^N,
\\
\tilde{(B)'} && (V^{(1)}) ^N =  c^N .
\eea
These are $N$ copies of 
flat planes in the flat $(W,V)$ space as is expected.
Similarly, all the other curves in the class considered in the previous section
are reduced to flat curves in this region.
Note that each of these curves is a paraboloid if seen in $(r_1,r_2,r_3)$.
For example, the curve $\tilde{(A)}$ is
\be
r_1 = - \frac{r_2^2+r_3^2}{2a} +\frac {a}{2}
\label{eq:paraboloid}
\ee
where $a=|c|^2 / N\! R$.

The supergravity solution realizing one of these curves $\Sigma_0$ as 
an M5-brane of world-volume $R^{1,3}\times \Sigma_0$ can be obtained as 
the same way as in ref.~\cite{AH}.
Also, the explicit forms of four-form field strength $F_{[4]}$ 
and the current $J_{[5]}^{(11)}$ of M5-brane can be given \cite{GM}.

Note that although for any flat plane in the space $(W,V)$
we can obtain the corresponding supergravity solution,
we limit ourselves to the class of flat curves such that they are represented as the
$r/N\! R \rightarrow 0$ limit of the curves described in the last section.
For example, curves like $(a W +b V)^N =c^N $ or $(a W +b \overline{V})^N =c^N$
are excluded unless $a=0$ or $b=0$.

Now, we give the solution of M5-brane corresponding to 
the curve $\tilde{(A)}$, $W^N = c^N$, explicitly. 
The metric is given by
\be
ds_{11}^2 = f_{5}^{-\frac {1}{ 3}} (\eta_{\mu\nu} dx ^{\mu}  dx ^\nu 
+dV d\overline{V})
+ f_{5}^{\frac {2}{ 3}}(dy^m dy^m + dW d\overline{W}) 
\label{eq:ds0r0}
\ee
where $\mu, \nu =0,\cdots 3$, $m=4,5,6$, and 
\be
f_5=1+\sum_{l=1}^{N}\frac{k}{(s^2+|W-c\,e^{\frac{2\pi l}{N}i}|^2)^{3/2}},
\quad s^2= y_4^2+y_5^2+y_6^2 .
\ee
Four-form field strength $F_{[4]}$ is
\be
F_{p_1p_2p_3p_4} =3k\, \epsilon_{ p_1p_2p_3p_4q}\,
\sum_{l=1}^{N}
\frac{\tilde{y^q}}{ (s^2+|W- c \,e^{\frac{2\pi l}{N}i}|^2 )^{5/2}}
\ee
where $\tilde{y}^p =(x_4,x_5,x_6,z_7,z_{10})$.
The Hodge dual of current associated with the M5-brane is given by
\bea
dF_{[4]} & \equiv &  \tilde{*}J^{(11)}_{[5]}\nonumber\\
&=&  \sum_{l=1}^{N}3k \Omega_4  \delta(y_4) \delta(y_5) \delta(y_6) 
\delta(W- c\,e^{\frac{2\pi l}{N}i}) 
dy_4\wedge dy_5\wedge dy_6\wedge dz_{7}\wedge  dz_{10}  . 
\label{eq:11jnc}
\eea
where $\tilde{*}$ denotes the 11-dimensional Hodge dual.
Similarly, we can give the same information as above for other M5-brane systems.

At this point, we comment on the definition of current with an M5-brane in the
background of KK-monopoles besides the near core region.
In the generic $r$:finite region, we do not have the supergravity solution and the definite 
form of the current cannot be obtained.
However, only from the knowledge of the location of the M5-brane
in a certain coordinate system of $M_{TN}$, e.g., $(\vec{r},x_{11})$, 
an approximate form of the Hodge dual of current associated with the M5-brane 
is determined.
That is, for a two dimensional curve described by
$g_i(\vec{r},x_{11})=0$ ($i=1,2$), Hodge dual of the current 
of the M5-brane at $ g_i=0$, $y_4=y_5=y_6=0$ is 
denoted by the 5-form which has the nonzero value only on the M5-brane:
\be
\tilde{*}J^{(11)}_{[5]}  \sim \delta(y_4) \delta(y_5) \delta(y_6)
 \delta(g_1) \delta(g_2)
 dy_4\wedge dy_5\wedge dy_6\wedge dg_1 \wedge dg_2 \, .
\label{eq:11dcform}
\ee
We will analyze the brane identification in 10 dimensions approximately
from the knowledge of the form of the wedge product 
$ dg_1 \wedge dg_2 $, in particular, the behavior of the term 
including $dx_{11}$.

\section{Ten-dimensional analysis}
We investigate the compactification of the systems of an M5-brane 
in the KK-monopole background.
We fix the compactification direction along $x_{11}$.
It is known that an M5-brane becomes an NS5-brane, a D4-brane or a bound state
of these two types of branes upon compactification.
However it is argued in refs.~\cite{AH,GM} that there is a sort of subtlety in
identification of branes when the original 11-dimensional system has 
M5-branes and  KK-monopoles together.
We clarify the problem by using the exact form of metric and four-form field strength in
the $r \rightarrow 0$ region and by performing the compactification explicitly.

Assuming the isometry along $x_{11}$ direction,
11-dimensional theory is related to 10-dimensions as follows:
\bea
ds_{11}^2& =& e^{-\frac{2 }{3}\Phi}\, ds_{10}^2 + e^{\frac{4 }{3}\Phi}
\,(dx_{11}+A_{[1]})^2
\label{eq:compactify}
\\
F_{[4]}^{(11)}&=& G_{[4]} + G_{[3]}\wedge dx^{11}. 
\label{eq:f4g43}
\eea
Here $ ds_{10}^2$ is 10-dimensional string metric, $\Phi $ is 10-dimensional 
dilaton and $ A_{[1]}$ is Ramond-Ramond 1-form potential.
As for the four-form field strength, we used the notation 
$G_{[4]}\equiv d B_{[3]}$ and $G_{[3]}\equiv d B_{[2]}$
where $B_{[3]}$ and $B_{[2]}$ are Ramond-Ramond 3-form potential
and NS-NS 2-form potential respectively.
There is another description of 10-dimensional 
four-form field strength $\tilde{G}_{[4]}$ given by
\be
F_{[4]}^{11}= \tilde{G}_{[4]} + G_{[3]}\wedge (dx^{11}+A_{[1]}) 
\ee
where $\tilde{G}_{[4]}= G_{[4]} - G_{[3]}\wedge A_{[1]}$.
This definition may be more convenient since $A_{[1]}$ and $G_{[4]}$
couple with each other and the related term in 10-dimensional supergravity action is
collected by the form $\sim \tilde{G}_{[4]}^2$ after all.
There is another advantage in using the latter definition \cite{GM}
which is related to the `gauge invariance.'
In terms of 11-dimensions, the gauge symmetry is represented as
a coordinate transformation $(\vec{r},x_{11})\rightarrow (\vec{r},x_{11}')$
where $x_{11}' = x_{11} +\gamma(\vec{r})$.
As a metric of the form eq.(\ref{eq:compactify}), it is written as gauge 
transformation of $A_{[1]}$ :
\be
\left\{
\begin{array}{rl}
dx_{11} &\rightarrow  dx_{11}'=dx_{11} +d \gamma 
\\
A_{[1]} &\rightarrow  A_{[1]}' = A_{[1]} - d \gamma .
\end{array}
\right.
\label{eq:gaugeinvxa}
\ee
It is easily seen that under this gauge transformation, 
$\tilde{G}_{[4]}$ is invariant but $G_{[4]}$ is not.
We will argue this point later again.

We will apply the general formula of compactification to our configurations
given in the last section.
In order to compactify the system along $x_{11}$, there should be an isometry along 
the direction.
Thus for any of our M5-brane systems with parameter $c\neq 0$,
we have to put infinite number of M5-branes uniformly 
along the $x_{11}$ direction, {\it i.e.}, we consider a family of curves with
$\alpha = 2\pi R \cdot k/n $ ($k=1,2,\cdots n$), and take the limit 
$n \rightarrow \infty $.
Note that if $c=0$, there is already an isometry along $x_{11}$.
For example, for the M5-brane system given by the curve $W^N=c^N$ 
in eq.(\ref{eq:ds0r0}), 
we should put infinite M5-branes on the circle of radius $|W| = | c|$,
which corresponds to take  
\be
f_5=1 +k' \,\int^{2\pi}_{0}
(s^2+(W-|c|e^{i \xi})^2)^{-\frac{3}{2}} \,d\xi 
\ee
where $k'$ is some regularized parameter.
Then, the compactification can be performed, and 
$ ds_{10}^2$, $\Phi $ and $ A_{[1]}$ are calculated according to 
eq.(\ref{eq:compactify}).

Now we define the currents associated with D4-branes and with NS-NS 5-branes
in 10-dimensions.
The most straightforward definition is 
\be
dG_{[3]} = *j_{[5]}\, ,\qquad dG_{[4]} = *j_{[4]}
\ee
where $*j_{[5]}$ and $*j_{[4]}$ are the 10-dimensional
Hodge dual of the currents 
of NS-NS 5-branes and D4-branes respectively.
The relation of these currents with 11-dimensional current $J^{(11)}_{[5]}$
is obtained by differentiating eq.(\ref{eq:f4g43}) as
\be
\tilde{*}J^{(11)}_{[5]} = *j_{[4]} + *j_{[5]} \wedge dx_{11}.  
\label{eq:j1145}
\ee
In ref.\cite{GM}, another definition of D4-brane current $\tilde{j}_{[4]}$ is
proposed as
\be
*\tilde{j}_{[4]} = d\tilde{G}_{[4]} - F_{[2]}\wedge G_{[3]}
\ee
where $F_{[2]} = dA_{[1]}$.
Two different definitions of D4-brane current coincide with each other when 
there is no R-R 1-form potential.
The main difference is that the latter definition,
$\tilde{j}_{[4]}$, respects the gauge invariance eq.(\ref{eq:gaugeinvxa})
of $A_{[1]}$ but the former does not, and that
the former assures the conservation of charge but the latter does not.
This means that the D4-brane charge defined by the current $\tilde{j}_{[4]}$
remains unchanged 
if we change the compactification direction as $x_{11}\rightarrow x_{11}'$.
In ref.\cite{GM}, 
by taking the current $\tilde{j}_{[4]}$ as physically relevant definition of D4-branes, 
charge non-conservation and brane transmutation are argued.

On the contrary, here we choose a standpoint of interpreting the naive
 definition of
D4-brane current $j_{[4]}$ as physically relevant one.
This viewpoint gives the physical meaning 
to the Dirac string type singularity associated with $A_{[1]}$.
It seems strange at first sight, however, we can offer
a reasonable explanation.
First of all, we remember that the gauge invariance 
of $A_{[1]}$, whose determination specifies the location of the singularity,
is not a symmetry of the compactified 10-dimensional 
theory in itself, but a symmetry in 11-dimensions as in eqs.(\ref{eq:gaugeinvxa}).
In compactifying to 10-dimensions, we have to specify $x_{11}$ or $x_{11}'$ definitely
as an eleventh direction along which the reduction is executed.
It is nothing but fixing of a gauge in terms of (\ref{eq:gaugeinvxa}).
The Dirac string singularity in 10-dimensions cannot be moved to another place
without changing the compactification direction.
Thus, looking in 10-dimensions, this singularity is not necessary to
be unphysical at least concerning the identification of branes.
In practice, in our generic configurations,
we will see that D4-branes come up from the 
singularity and go to infinity along an NS5-brane. 

Before demonstrating the consequence of the above interpretation
of 10-dimensional currents, in the next section
we deal with similar configurations in which there exists Dirac string type singularity. 
Using the models, we
examine the relation between the singularity and the brane identification.

\section{M-branes in the stringy cosmic string backgrounds}
We take the stringy cosmic string solution as 11-dimensional supergravity background.
The solution is known to represent a configuration with some
parallel D7-branes ($[1,0]$-branes) or their SL$(2,{\bf Z})$
dual $[p,q]$7-branes if it is taken as a IIB background~\cite{Va}. 
It is also interpreted as the compactification of 12-dimensional
F-theory on K3, or non-compact K3, which admits elliptic fibration.
We will only deal with non-compact K3 manifolds,
especially with at most one point-like singularity of $A_N$ type as a total space.
We first explain the corresponding IIB background, and then,
by using the conjectured duality between F/(K3$\times S^1$) and M/K3,
or IIB/$S^1$ and M/$T^2$,
we construct the corresponding background in terms of M-theory.

It is important that the complex modulus $\tau $ of the fiber torus 
of the non-compact K3 is only determined up to SL$(2,{\bf Z})$ as a function
on the base manifold which is isomorphic to some orbifold of ${\bf C}$.
This SL$(2,{\bf Z})$ is the IIB S-duality group and parametrized as 
$\tau = \chi + ie^{-\phi}$ where $\chi$ is axion and $\phi$ is dilaton field.
If we try to assign some definite value of $\tau $ at each point in the base manifold,
there should exist branch cut
coming up from each 7-brane to infinity \cite{GZ}. 
The behavior of the $(p,q)$-string, or some extended configuration 
including three string junctions,
in the 7-brane background has been investigated in various situations.
Here a $(p,q)$-string is a bound state of $p$ fundamental strings and 
$q$ D-strings and can end on a $[p,q]$7-brane.

\begin{figure}[hbt]
\begin{center}
\includegraphics[width=11cm,clip]{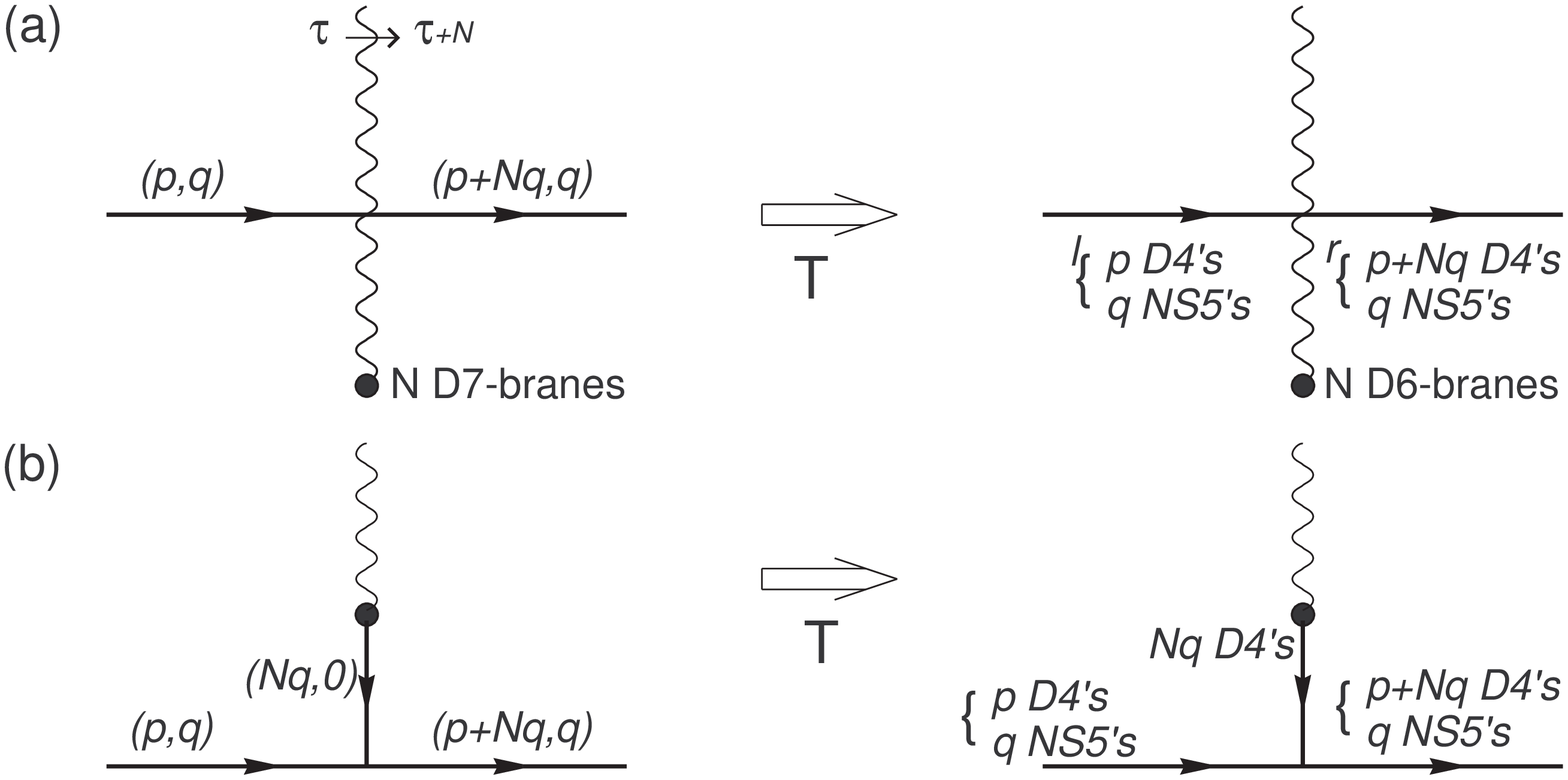}
    \end{center}
    \caption{\small $(p,q)$5-branes in the background of $N$ coincident D7-branes
seen in the plane transverse to the D7-branes.
The charge assignment of 5-branes changes if the branes cross the branch 
cut.
Brane creation from D7-branes are explained in terms of charge conservation 
of (a) and (b).
T-dualized IIA theory picture of this configuration is depicted in the right hand side of each figure.}
    \label{fig:pq7}
  \end{figure}

Here we focus on the role of branch cut and take some duality transformation
to obtain the M-theory counterpart. 
Moreover, since we want M5-branes in the dual 11-dimensional theory,
we apply the argument to the case of $(p,q)$5-branes of world-volume 
$R^{1,4}\times l$ where $l$ is a one dimensional object in the base space.
Here a bound state of $p$ D5-branes and $q$ NS5-branes is denoted 
by a $(p,q)$5-brane. 
If a $(p,q)$5-brane goes around $N$ D7-branes counterclockwise,
the brane is converted into the $(p+Nq,q)$5-brane 
with the shift $\tau \rightarrow \tau +N$ as in Fig.{\ref{fig:pq7}}.
This means that if the 5-brane crosses the branch cut singularity, then 
the charge assignment of the brane changes as in Fig.{\ref{fig:pq7}},
and if it crosses $N$ 7-branes, $Nq$ new D5-branes are created from the 7-branes 
to form a three 5-brane junction \cite{AhHa}. 
This is regarded as U-dual of original Hanany-Witten effect where
a D3-brane is created if D5-brane crosses an NS5-brane \cite{HW}. 

Using the duality conjecture between F-theory and M-theory,
we can obtain the M-theory background from this 7-brane background.
Consequently, the fiber torus turns up as 
 a geometrical object in 11-dimensions, {\it i.e.}, 
the background becomes elliptic (non-compact) K3 manifold times 7-dimensional
Minkowski space.
There is another method to reach the M-theory background;
beginning with the explicit background of 7-branes,
we perform T-duality along the appropriate world component of 7-branes 
as well as $(p,q)$5-branes,
and then go up to 11-dimensions \cite{KS}. 
The $(p,q)$5-brane turns into a bound state of $p$ D4-branes and 
$q$ NS5-branes in IIA theory, and then becomes an M5-brane winding around
$(p,q)$-cycle of the torus.
In particular, if we begin with $N$ D7-branes, 
we obtain the IIA background with $N$ D6-branes
and thus we see that this background resembles
the compactification of KK-monopole background in the sense that both
 deal with D6-branes.
However, unlike the KK-monopole case, the number $N$ must be less than 24
for geometrical reason.

Since the supergravity solutions including both 7-branes and 
$(p,q)$5-branes are not known, we study the configuration by
putting 5-branes as probes in the background of 7-branes.
Since this configuration reserves some supersymmetry,
the M5-brane transformed from IIB $(p,q)$5-brane
must be holomorphically embedded in the corresponding M-theory background \cite{BBS}.
We use the same embedding as in the case of M2-branes
given in ref.\cite{KS}. 

For the background with $N$ D7-branes at the origin of the base space 
spanned by $(z,\bar{z})$,
the metric in terms of dual 11-dimensional theory is
\be
ds^2_{scs}= \eta_{\mu \nu}dx^{\mu}dx^{\nu} + 
e^{\Phi(z,\bar{z})}dz d\bar{z} + \tau_2 ^{-1} d \zeta d\bar{\zeta} 
\ee
where
\be
e^{\Phi(z,\bar{z})} = \tau_2\eta^2\bar{\eta}^2| z|^{-\frac{N}{6}},
\ee
$\tau = \tau_1 + i \tau_2 $ and 
$\zeta = \tilde{u}+\tau \tilde{v}$.
The torus is spanned by the periodic coordinates 
$(\tilde{u},\tilde{v})\sim (\tilde{u}+2\pi R, \tilde{v}+2\pi R)$.
And the curve wound along $(p,q)$-cycle of the torus is represented as 
$q\tilde{u}-p\tilde{v} = const.$.
The modulus $\tau$ is some holomorphic function of $z$, 
and the behavior in the $z \rightarrow 0$ limit is 
\be
\tau(z)\sim \frac{N}{2\pi i} \log z ,
\ee
which requires a branch cut from the origin.
As we mentioned earlier, this branch cut is coordinate singularity in 11-dimensions.
Other notations are the same as ref.\cite{Va,MA}.
By compactifying this metric along $\tilde{u}$ direction, 
we obtain the IIA background of D6-branes with
\bea
e^{\frac{4}{3} \Phi} &=& \tau_2^{-1},
\\
A_{[1]} &=& \tau_1 d\tilde{v} \stackrel{z\sim 0}{\to} 
\frac{N}{2\pi} \arg z \, d\tilde{v} .
\eea
We see that the coordinate singularity is now interpreted as the Dirac string
singularity with respect to one-form potential $A_{[1]}$ in IIA theory.
In the three-dimensional space spanned by $(z,\bar{z},\tilde{v})$,
the singularity is two-dimensional, which is different from the case of 
KK-monopole solution.

Now in this background we put an M5-brane of world-volume
 $R^{1,3}\times \Sigma $ where $\Sigma $ is a holomorphic curve 
winding around the $(p,q)$-cycle of the torus.
We choose the complex structure such that it is orthogonal to that defined by 
$(z,\zeta)$ \cite{KS}.
If we use the discussion in the end of section 5, 
the Hodge dual of the current associated with this M5-brane is given as
\be
\tilde{*}J^{(11)}_{[5]} \sim
\delta(y_4) \delta(y_5) \delta(y_6)
 \delta(q \tilde{u}-p \tilde{v}) \delta(g)
 dy_4\wedge dy_5\wedge dy_6\wedge d(q \tilde{u}-p \tilde{v}) \wedge dg .
\ee
Here $g=g(z, \bar{z})$ is defined as a real one-dimensional
line in the $z$-plane representing the location of branes in the plane
\cite{GZ}.
If the brane is located over the both sides of the singularity as in Fig.2(a),
coordinates $(\tilde{u},\tilde{v})$ are discontinuous at the singularity, 
and the definition of $(p,q)$-cycle changes at the singularity:
If we cross the singularity counterclockwise, $(p,q)$-cycle changes to 
$(p+Nq,q)$-cycle.

In the flat spacetime $ {\cal M}^{1,8}\times T^2$,
an M5-brane winding along $(p,q)$-cycle of the torus
leads to the bound state of $p$ D4-branes and $q$ NS5-branes
upon compactification along the $p$-cycle ($\tilde{u}$ direction). 
This formula is also applicable to our case and the  
current associated with D4-brane charge is represented by the NS5-brane 
current  if $q\neq 0$ as 
\be
*j_{[4]} = - *j_{[5]}\wedge \frac{p}{q} d\tilde{v} 
\ee
where we used eq.(\ref{eq:j1145}).

If the M5-brane is put such that it crosses the singularity, the interpretation of 
D4-brane charge after compactification depends on the place of the brane, 
left or right of the singularity. 
Note that the current associated with NS5-brane is determined 
independently of the location of the singularity.
In the case of the brane depicted in Fig.2(a), 
D4-brane current is given as
\be
\left\{
\begin{array}{rl}
*j_{[4]}^l &= - *j_{[5]}\wedge \frac{p}{q} d\tilde{v} ,
\\
*j_{[4]}^r &= - *j_{[5]}\wedge \left( \frac{p}{q} +N \right)d\tilde{v}
\end{array}
\right.
\ee
where the superscript $l$ and $r$ denote left and right hand side of 
the singularity respectively as in Fig.2(a).

{}From this assignment of currents, it can be seen 
as if new D4-brane charges are created from the string-like
singularity, though
there is no discontinuity in the original 11-dimensional viewpoint.
This is not so peculiar since
if we perform T-duality transformation along $q$-cycle ($\tilde{v}$ direction),
 the determination of current is consistent with that
of $(p,q)$5-branes in terms of the SL$(2,{\bf Z})$ symmetry of IIB 
background with 7-branes.

\section{Definition of D4-brane charges : Identification of D4-branes}
We know from the discussion of M5-branes in 
the stringy cosmic string background where the location of the 
Dirac string type singularity plays an important role in identifying the brane as 
an NS5-brane, a D4-brane, or a bound state of them
in the compactified 10-dimensions.
Now we return to the case of KK-monopole background and consider
the identification of 10-dimensional brane dimensional reduced from an M5-brane.

\begin{figure}[hbt]
\begin{center}
\includegraphics[width=13.5cm,clip]{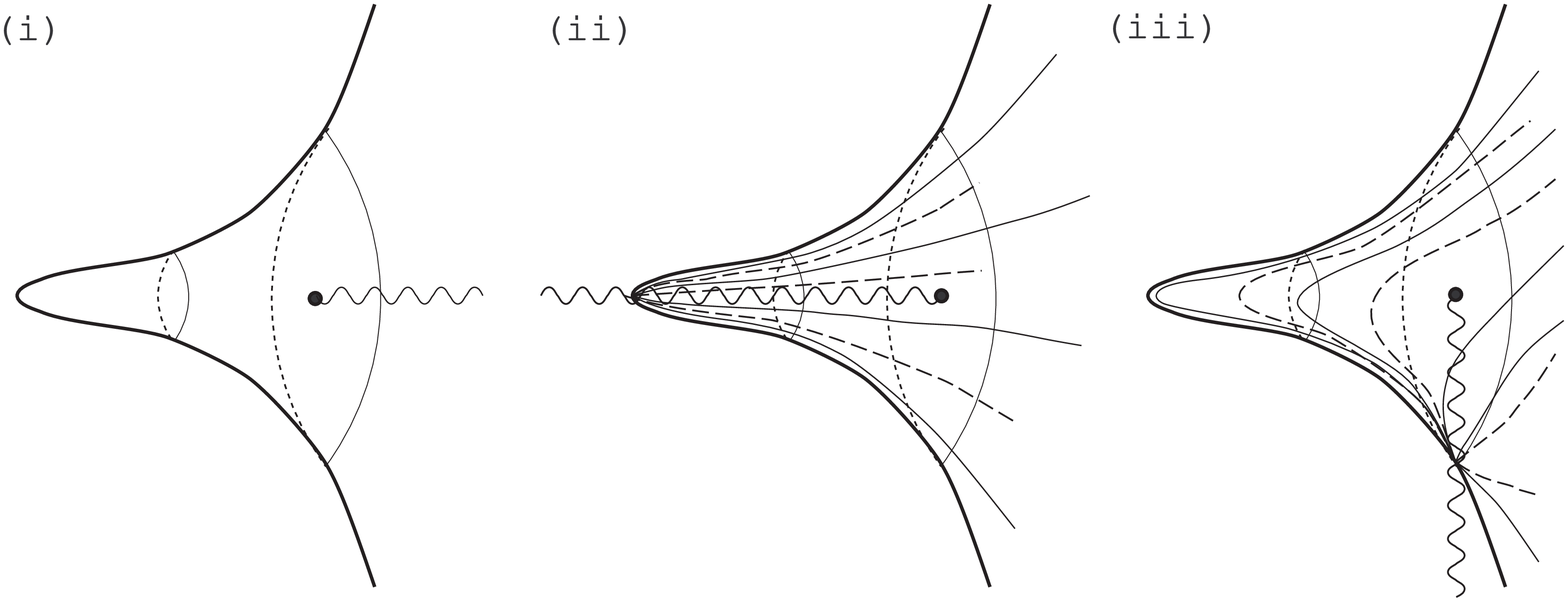}
    \end{center}
\caption{\small The appearance of D4-branes on the NS5-brane is depicted by 
the striped pattern for three representative cases
 eqs.(\ref{eq:f1})$\sim$(\ref{eq:f3}).
D4-branes arise at the singularity depicted by the wavy lines and
go to infinity along the NS5-brane.
}
    \label{fig:curve123}
  \end{figure}

We explicitly argue three types of M5-branes specified by the 
following curves in $M_{TN}$ among others : 
\bea
(i)&  &  f_1 \equiv|w_0|-e^{-\frac{b}{N\! R}}=0 \qquad [\Leftrightarrow (A)]
\label{eq:f1}
\\
(ii) &  &  f_2 \equiv \left| \frac{v_0}{w_0}\right| - e^{-\frac{b}{N\! R}}=0 \qquad 
[\Leftrightarrow(B)]
\label{eq:f2}
\\
(iii)&  & f_3 \equiv | {w_1}|- e^{-\frac{b}{N\! R}}=0 \qquad 
[\Leftrightarrow (A)']
\label{eq:f3}
\eea 
They are the typical curves concerning the location of 
the singularity relative to the curve seen in $(r_1,r_2,r_3)$ space 
(see Fig.1 or \ref{fig:curve123}).
The curve $(i)$ does not intersect the singularity for any value of $b$,
$(ii)$ intersects the singularity at the vertex of the curve for any $b$
and $(iii)$ intersects the singularity asymmetrically.
Note that all other curves belonging to our family are classified 
into the same class as $(iii)$ in the sense that they intersect the 
singularity asymmetrically. 
Also notice that this class is further divided into two classes: One class [$(iii)$A]
consists of curves which cross the singularity only one time for all values of $b$, and 
the other [$(iii)$-2] is the set of curves that do not cross the singularity 
if we take $b$ small enough.
The curve $(iii)$ in (\ref{eq:f3}) itself belongs to the former class $(iii)$-1.
A curve in the class $(iii)$-2 can be realized by choosing a complex structure,
e.g., given by the holomorphic coordinates $(w_0+w_1, v_0+v_1)$.
In the limit $b/N\! R \rightarrow \infty$, the point of intersection becomes closer to 
the D6-branes for all curves.
Note that the curves in the class $(iii)$-2 has another intersection point 
at large $r$ in the limit.

We analyze the configuration of D4-branes and NS5-branes  
by the currents $j_{[4]}$ and $j_{[5]}$ in a similar way as in the previous section.
First we consider the 11-dimensional current $J_{[5]}^{(11)}$ for each of three curves.
They are obtained by applying the general formula eq.(\ref{eq:11dcform}) to the above 
cases :
\bea
(i) &  &  \tilde{*}J_{[5]}^{(11)} \sim  
\delta(y_4) \delta(y_5) \delta(y_6) \delta(f_1(\vec{r})) 
dy_4 \wedge dy_5 \wedge dy_6 \wedge df_1 \wedge dx_{11}  ,
\\
(ii)&  &  \tilde{*}J_{[5]}^{(11)} \sim  
\delta(y_4) \delta(y_5) \delta(y_6) \delta(f_2(\vec{r})) 
dy_4 \wedge dy_5 \wedge dy_6 \wedge df_2
\wedge (dx_{11}-N\! R d\psi)  ,
\\
(iii)&  & \tilde{*}J_{[5]}^{(11)} \sim  
\delta(y_4) \delta(y_5) \delta(y_6) \delta(f_2(\vec{r})) 
dy_4 \wedge dy_5 \wedge dy_6 \wedge df_2
\wedge (dx_{11}- N\! R d \chi)
\eea
where $f_i$ is given in eqs.(\ref{eq:f1})$\sim$ (\ref{eq:f3}) and 
$\chi$ is in eq.(\ref{eq:dx111}).
Note that in the near core limit, we can show the definite form of the 11-dimensional
current as in eq.(\ref{eq:11jnc}).
In the reduced 10-dimensional theory,
the relation between D4-brane current $j_{[4]}$ and NS5-brane current $j_{[5]}$
for each brane is determined independent of the scale of $r$ as
\bea
(i)&  &  *j_{[4]} = 0  , 
\\
(ii)&  &  *j_{[4]} = *j_{[5]}\wedge (-N\! R d\psi)  ,
\\
(iii)&  &  *j_{[4]} = *j_{[5]} \wedge (-N\! R d \chi) \,  .
\eea
This means that there is no D4-brane current in the case $(i)$. 
In other cases, D4-branes can be present only on an NS5-brane.
By investigating the form of $*j_{[4]}$ on the NS5-brane,
we see how D4-branes are stretched on the NS5-brane.
If the NS5-brane intersects with the singularity only once, 
then it is interpreted that 
the smeared D4-branes come up from the Dirac singularity, 
{\it i.e.}, the D4-branes are created from the singularity, and go to infinity. 
For the curves in the class $(iii)$-2, 
there are two points on the corresponding NS5-brane 
where the D4-branes can be created or absorbed.
Note that since $* j_{[4]}$ is proportional to the number $N$,
the number of D4-branes smeared on the NS5-brane
is also proportional to $N$.
The results are depicted in Fig.\ref{fig:curve123}.


\section{On the brane creation phenomena}
Brane creation was first discussed in the flat IIB background \cite{HW}
as the Hanany-Witten effect representing a phenomenon 
that a new brane is created by crossing of certain two types of branes.
This effect is confirmed by charge conservation, although the process of brane creation
has not yet been clarified in the framework of supergravity theory. 
One reason for this is that it is difficult to describe the situation of a
brane ending on another brane as a solution of supergravity.
Now, our discussion given in the preceding sections enables us to 
study the process of brane creation in the vicinity of D6-branes
in terms of the corresponding exact solution of supergravity.

Using the argument, we now discuss how the brane creation is explained in the near core
region of KK-monopoles and also consider the extension to the $r$:finite region of 
$M_{TN}$.
In particular, we again deal with three types of M5-branes $(i)\sim (iii)$
and their compactification.

First, consider the curves $(ii)$ and $(iii)$.
They intersect the Dirac string singularity if the parameter $b$ is taken
 to be large enough.
As was indicated in the last section, it is interpreted in 10-dimensions that 
D4-branes come up from the point of intersection with the singularity
and go to infinity or another intersection point along the NS5-brane.
In the near core limit $r/N\! R \rightarrow 0$, 
each curve is a paraboloid and the shape of the paraboloid changes with the parameter
 $|c|$ as in eq.(\ref{eq:paraboloid}).
In the limit $c/N\! R \rightarrow 0$ ({\it i.e.}, $b/N\! R \rightarrow \infty$),
 the paraboloid becomes like a thin tube and degenerates to an object like a half-string.
As $c$ approaches $0$, the NS5-brane bends so as to wrap the D6-branes and 
gives multipole moment if measured far away from the brane \cite{GM}.
Finally in the limit $c \rightarrow 0$, the total NS5-brane charge vanishes.
On the other hand, in this limit, all the D4-branes spread over the NS5-brane 
gather to form a bundle of coincident D4-branes coming up from D6-branes.
This process of disappearing the NS5-brane and the assembling of D4-branes 
into one place may be regarded as the brane creation in the near core limit.
  \begin{figure}[hbt]
\begin{center}
\includegraphics[width=10cm,clip]{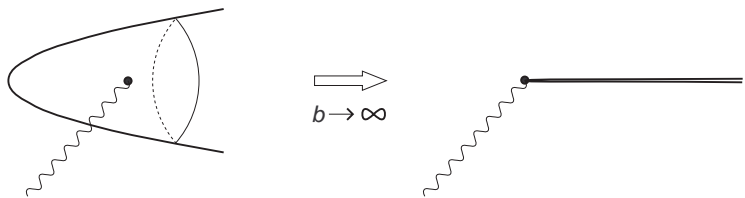}
    \end{center}
\caption{\small In the limit $b/N\! R \rightarrow \infty$ ($c/N\! R \rightarrow 0$), 
the NS5-brane charge disappears and 
only the coincident D4-branes coming up from D6-branes remain 
in the near core region.
This is the brane creation process of D4-branes for the M5-branes of the classes
$(ii)$ and $(iii)$.}
  \end{figure}

Extending this process of brane creation to the outer region, $r$:finite region,
we see a transition of branes with respect to the value $b$. 
For the curves $(ii)$ and $(iii)$-1, 
the transition process is essentially the same as in the near core region, 
since these curves already intersect with the singularity in the $b/N\! R \rightarrow -\infty$.
A curve in the class $(iii)$-2 begins with a pure NS5-brane 
if $b$ is small enough.
Then as $b/N\! R$ becomes larger, the NS5-brane begins to bend and at some 
value of $b$ the curve comes in contact with the singularity, and after that 
we have two points of intersection and the D4-brane charges turn out.
Note that in the limit $b/N\! R \rightarrow \infty$,
NS5-brane charge disappears and only D4-brane charge remains roughly in the
region $r$ smaller than $b$ for these cases.
  \begin{figure}[hbt]
\begin{center}
\includegraphics[width=14.5cm,clip]{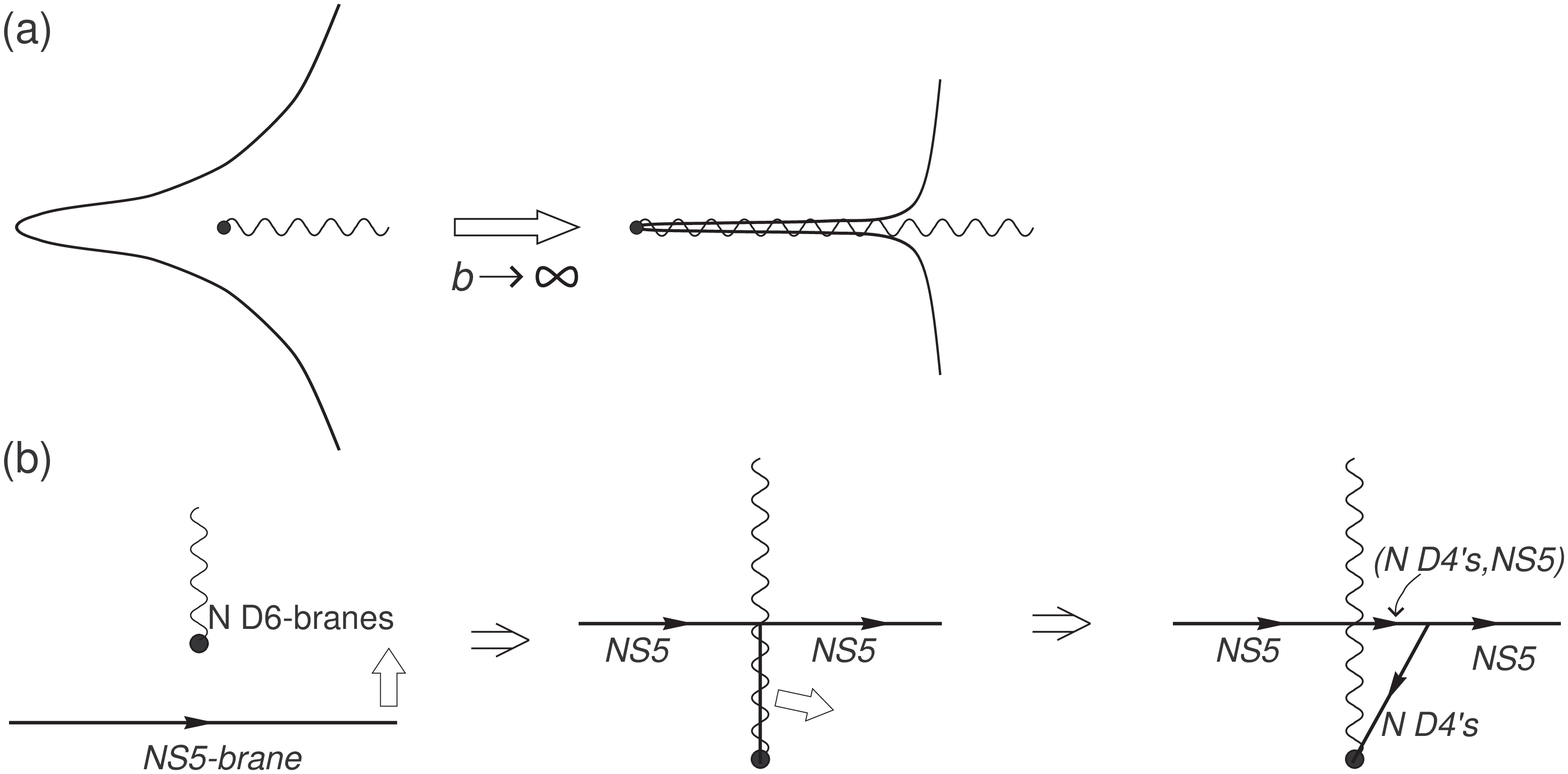}
    \end{center}
\caption{\small (a) The $b \rightarrow \infty$ limit of the NS5-brane specified by
the curce $(i)$. There is no D4-brane charges, though a D4-brane like object
appears in the limit.
(b) Similar phenomenon occurs in the case of stringy cosmic string background:
When an NS5-brane crosses $N$ D6-branes, D4-branes which should be created
from D6-branes cannot be seen if the branch cut lies on the D4-branes. 
If we shift the cut, D4-branes can be observed.
 }
    \label{fig:branecreation}
  \end{figure}

On the other hand, consider the curve $(i)$ for which there is no D4-brane 
charge on the NS5-brane.
In the limit $b/N\! R \rightarrow \infty $, the brane 
is bent as the same way as other cases, however, 
we have to take another explanation as others.
In particular, in the region near the D6-branes,
this curve reduces to an object which has an appearance of a D4-brane ending on D6-branes,
but has neither NS5-brane charges nor D4-brane charges.
We have a problem how to interpret the object.
The resolution of this can be done by noticing the singularity again:
The singularity is placed as overlapped on the D4-brane-like object
in the limit $b/N\! R \rightarrow \infty $. 
Thus the situation is exceptional and the identification of 
the brane-like object cannot be performed.

The analogous structure is found in the IIB stringy cosmic string background
where some 5-branes ending on $N$ coincident D7-branes
are exactly on the branch cut.
In terms of IIA theory, the branes 
must be identified as D4-branes, while it cannot be seen from charge conservation.
The identification can be done by shifting the branes away from the branch cut
singularity as in Fig.\ref{fig:branecreation}(b).

In our case of KK-monopole background, shifting the D4-brane 
away from the singularity 
corresponds to taking another curves holomorphic with respect to different
 complex structure, or shifting the compactification direction as in 
eq.(\ref{eq:gaugeinvxa}).
By using either method, the identification of the branes can be done.
This resolves the puzzle we stated.

\section{Summary and Discussion}
We have proposed a method of identifying the branes in the IIA background especially
in the presence of D6-branes obtained by compactification of 11-dimensional 
Kaluza-Klein monopole solution.
It is essentially the same as the identification of branes in the IIB
background with 7-branes.
We have also discussed the brane creation from D6-branes.
In particular, since we know exact supergravity solutions of M5-branes 
in the near core region of KK-monopoles in 11-dimensions, 
we have clarified the mechanism of brane creation in this region
as a compactified 10-dimensional theory. 

Now we explain the relation between the brane creation based on our supergravity 
argument and the original Hanany-Witten argument based on flat branes
in the flat background.
In the flat space argument, created branes between two branes have finite length. 
On the other hand, in our argument 
the created branes are interpreted as half-infinite branes:
The D4-branes coming up from the D6-branes are not cut on the NS5-brane,
but continue along the NS5-brane.
The difference may be related to the fact that there exist Dirac string singularities in the
background with D6-branes.
If we could obtain supergravity solutions of 
the original Hanany-Witten configurations, 
we would be able to clarify whether a finite brane can exist.

We comment on the case of M2-branes instead of M5-branes.
In this case, we have to take care of the definition of currents after compactification:
In order to proceed with the argument on identification of F-strings or D2-branes
as the same method as in the case of M5-branes,
it seems to be necessary to use the definition of 10-dimensional field strength as
$\tilde{*}F_{[4]}^{(11)}= *\hat{G}_{[3]} + *\hat{G}_{[4]}\wedge dx^{11}$
instead of $ G_{[3]}$ and $ G_{[4]}$ in eq.(\ref{eq:f4g43}).
Although the two definitions coincide with each other if there are no Kaluza-Klein
charges, in general $ G_{[n]} \neq \hat{G}_{[n]}$ in the presence of KK-charges.
Nevertheless we do not know direct reason to change the definition of 10-dimensional 
field strengths as above depending on the objects we deal with.
Even if we decide to use such a definition, the problem still remains in the situations 
where both M5-branes and M2-branes exist simultaneously.
(In such a situation, Chern-Simons term $F \wedge F \wedge A$ in 11-dimensional 
supergravity action may play an important role.)
Note that a same kind of confusion in defining the field strengths in 10-dimensions  
was pointed out in the previous argument \cite{AH,GM}.
We do not have definite explanation of this puzzle at the present point.

Moreover, note that our 10-dimensional configuration with Dirac string singularity is
based on singular compactification from the 11-dimensional background.
We do not know if such dimensional reduction is 
consistently described as a compactification of string theory.

\paragraph{Acknowledgments}
This work was supported in part by JSPS Research Fellowships for 
Young Scientists. 

\end{document}